\documentclass[journal,twocolumn]{IEEEtran}

\usepackage{amsfonts}
\usepackage{times}
\usepackage{graphicx}
\usepackage{epstopdf}
\usepackage{latexsym}
\usepackage{dsfont}
\usepackage{amssymb}
\usepackage{amsmath}
\usepackage{cite}
\usepackage{verbatim}

\newcommand{\figref}[1]{{Fig.}~\ref{#1}}

\def\bb0{{\mathbb{0}}}

\def\bb{{\mathbf{b}}}

\def\bff{{\mathbf{f}}}

\def\bh{{\mathbf{h}}}

\def\bp{{\mathbf{p}}}

\def\bs{{\mathbf{s}}}

\def\bx{{\mathbf{x}}}

\def\b0{{\mathbf{0}}}

\def\bbC{{\mathbb{C}}}

\def\bbE{{\mathbb{E}}}

\def\bbP{{\mathbb{P}}}

\def\bbR{{\mathbb{R}}}

\def\bbZ{{\mathbb{Z}}}

\def\cF{\mathcal{F}}

\def\sf0{{\mathsf{0}}}

\usepackage{graphicx}
\usepackage{epstopdf}
\usepackage{enumerate}
\usepackage{algorithm}
\usepackage{amsmath}
\usepackage{mathrsfs}
\usepackage{float}
\usepackage{color}
\usepackage{makeidx}
\usepackage{bm}
\usepackage{cleveref}
\usepackage{url}
\usepackage{steinmetz}
\usepackage{varwidth}
\usepackage{soul}
\usepackage{bbm}
\usepackage{empheq}
\usepackage{mathtools}
\usepackage{cite}
\usepackage{balance}

\def \rm {\mathrm}

\usepackage{amsfonts}
\usepackage{booktabs}
\usepackage{siunitx}

\begin{document}
\title{LiDAR Aided Future Beam Prediction in Real-World Millimeter Wave V2I  Communications}
\author{Shuaifeng Jiang, Gouranga Charan, and Ahmed Alkhateeb\\ \textit{School of Electrical, Computer and Energy Engineering - Arizona State University} \\ \textit{Emails: \{s.jiang, gcharan, alkhateeb\}@asu.edu}\thanks{This work is supported by the National Science Foundation under Grant No. 2048021.} }

\maketitle
\begin{abstract}
This paper presents the first large-scale real-world evaluation for using LiDAR data to guide the mmWave beam prediction task. A machine learning (ML) model that leverages the LiDAR sensory data to predict the current and future beams was developed. Based on the large-scale real-world dataset, DeepSense 6G, this model was evaluated in a vehicle-to-infrastructure communication scenario with highly-mobile vehicles. The experimental results show that the developed LiDAR-aided beam prediction and tracking model can predict the optimal beam in $95\%$ of the cases and with more than $90\%$ reduction in the beam training overhead. The LiDAR-aided beam tracking achieves comparable accuracy performance to a baseline solution that has perfect knowledge of the previous optimal beams, without requiring any knowledge about the previous optimal beam information and without any need for beam calibration. This highlights a promising solution for the critical beam alignment challenges in mmWave and terahertz communication systems. 
\end{abstract}

\begin{IEEEkeywords}
beam tracking, LiDAR, machine learning, DeepSense 6G, real-world data
\end{IEEEkeywords}

\section{Introduction}
To support the growing demand for data rate, millimeter-wave(mmWave) and terahertz (THz) communication systems have gained increasing interest \cite{heath2016overview}. Although the mmWave/THz systems have larger bandwidth, they suffer from the inceasing path-loss implications.  To achieve sufficient signal-to-noise ratio (SNR), these communication systems rely on massive antenna arrays and narrow beams. However, obtaining the optimal narrow beams often requires prohibitively large beam training overhead. For high mobility applications, this becomes more infeasible since the channel rapidly changes and beam training is more frequently performed. With this motivation, novel approaches that can reduce or even eliminate the beam training overhead are desired. Due to the propagation nature of mmWave/THz signals, the narrow beams are highly directional and the beam management problem particularly relies on the position/direction of the base station (BS) and the user equipment (UE), and the surrounding environment. To that end, the sensing information of the environment and the UE could be utilized to guide the beam management and reduce beam training overhead \cite{alkhateeb2018deep,alrabeiah2020deep,charan2021vision, demirhan2021radarbeam}.
\par
Prior works have studied improving mmWave/THz beam selection and blockage prediction based on a variety of sensing modalities \cite{alkhateeb2018deep,alrabeiah2020deep,charan2021vision, demirhan2021radarbeam}. One conventional direction is using wireless sensing information. In \cite{alkhateeb2018deep}, the author proposed to leverage wireless received signals as a unique signature for the UE position and its interaction with the surrounding environment. \cite{alrabeiah2020deep} exploits the sub-6 GHz channel, which is easier to obtain compared to the mmWave channel, to predict the mmWave beam and blockage status. Another direction gaining increasing interest is exploiting the vision/camera information. \cite{charan2021vision} leverages the camera signals to proactively predict dynamic link blockages for mmWave systems.  In \cite{demirhan2021radarbeam}, the authors proposed to use the radar signals to sense the UE to predict the 6G beam.
\par
However, each sensing modality mentioned above has its own drawbacks. The wireless received signal and the radar sensing signals occupy wireless resources. The performance of the camera/vision-based sensing degrades under poor lighting conditions, and it leads to privacy concerns. Extra signaling overhead is needed for the BS to obtain position information of the UEs, and the accurate position information can also cause privacy problems. This motivates us to investigate the other alternative: the light detection and ranging (LiDAR) sensors.
\par
In this paper, we propose to achieve mmWave/THz beam prediction and beam tracking relying on the LiDAR sensing data. Our contribution is summarized as follows.
\begin{itemize}
\item We propose to leverage LiDAR sensory data  to aid the mmWave beam prediction and tracking tasks. While using LiDAR to predict current beams has been previously investigated in \cite{8642397,9681382}, this is the first paper to investigate leveraging LiDAR data to also predict \textit{future} beams. .
\item We propose an efficient machine learning (ML) model for the LiDAR aided  beam prediction tasks.
\item To evaluate the performance of the proposed LiDAR beam prediction and tracking approach, this paper presents the first large-scale real-world evaluation results based on  a vehicle-to-infrastructure communication scenario of the DeepSense 6G dataset. 
\end{itemize}
Evaluation results show that the proposed LiDAR beam prediction and tracking approach can achieve high accuracy performance: it can predict the optimal beam in $95\%$ of the cases with beam training overhead reduced from $64$ to $5$. This demonstrates the potential of LiDAR sensing information in real-world mmWave/THz communication systems.
\par

\begin{figure*}[t]
	\centering
	\includegraphics[width=1\linewidth]{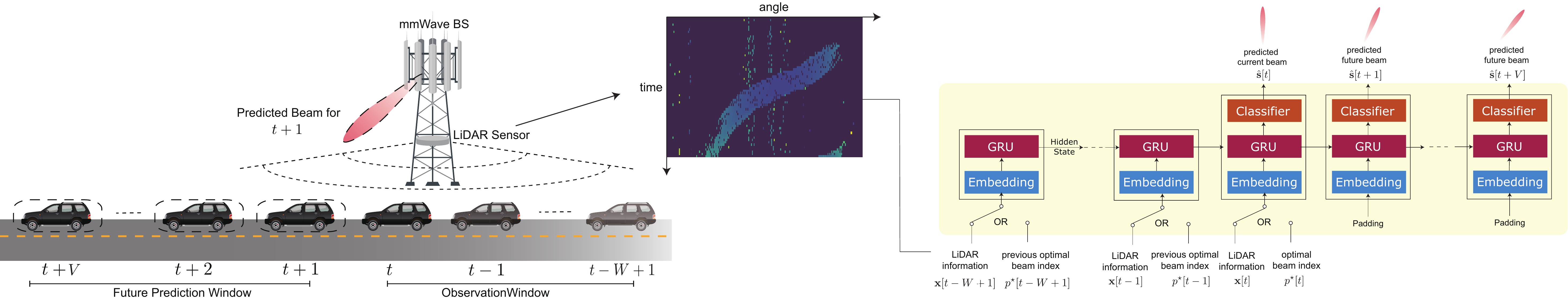}
	\caption{This figure illustrates the considered system model: the BS senses the environment and the moving UE with a LiDAR. The obtained sensing information is then utilized for the BS beam prediction and beam tracking. The figure also shows the block diagram of the proposed ML models. The ML model consists of the embedding block, the GRU feature extractor block, and the classifier block.}
	\label{fig:system_model}
\end{figure*}

\section{System Model and Problem Formulation}\label{System and Problem Formulation}
In this section, we first introduce the adopted system model. Then, we define the two considered beam management tasks, namely LiDAR aided current and future beam prediction.

\subsection{System Model}
As shown in Fig. \ref{fig:system_model}, we consider a communication system model where a base station (BS) serves a mobile user equipment (UE) over a mmWave frequency band. The BS has an antenna array of $N$ elements, and is also equipped with a LiDAR sensor that collects sensing measurements of the communication environment. For simplicity, the UE is assumed to be single-antenna. Let $\bh[t] \in \bbC^{N \times 1}$ denote the channel vector between the BS and the UE at time step $t$.
In the downlink transmission, if the BS sends a complex symbol $s\in \bbC$, the downlink received signal $y[t]$ at the UE can be written as
\begin{align}\label{eq:signal model}
y[t] = \bh^H[t] \bff[t] s + n[t],
\end{align}
where $\bff[t] \in \bbC^{N \times 1}$ is the transmit beamforming vector the BS used at time step $t$, and $n$ is the receive noise, satisfying $\bbE[n[t] n^H[t]] = \sigma_n^2$. The downlink signal $s$ satisfies $\bbE\left[ss^H\right]=P$ with $P$ denoting the transmit power. The BS is assumed to adopt a pre-defined beamsteering codebook ${\boldsymbol{\cF}} = \{\bff_1,\hdots, \bff_M\}$ of size $M$, \textit{i.e.}, $\bff[t] \in {\boldsymbol{\cF}}$.

\subsection{Problem Formulation and ML Task Definition}
Finding the optimal beam $\bff^\star$ (that maximizes the receive power) out of the codebook ${\boldsymbol{\cF}}$ typically requires beam training with high overhead that scales with the number of antennas \cite{heath2016overview}. In this paper, we investigate how to leverage the LiDAR sensory information to reduce this beam training overhead. In particular, we define and formulate two problems: (i) LiDAR-aided beam prediction, which predicts the current optimal beam with the LiDAR information obtained at the same time step. and (ii) LiDAR-aided \textit{beam tracking}, which predicts the future optimal beams using the LiDAR information obtained in the previous time steps. Since the two tasks target the prediction of the BS optimal beam at a particular time step, we start by defining the {optimal beam as the one maximizing the  receive power}, which is given by  
\begin{equation}\label{eq:bbest beam}
\bff^\star[t] =\underset{\bff[t] \in {\boldsymbol{\cF}}}{\arg\max}\ \big|\bh^H[t] \bff[t]\big|^2.
\end{equation}
Since the optimal beam $\bff^\star[t]$ is constrained by the pre-defined codebook ${\boldsymbol{\cF}}$, the goal of predicting the optimal beam is equivalent to predicting the index of the optimal beam. Let $p[t]\in\{1,\hdots,M\}$ denote the beam index at time step $t$, then we can write the optimal beam index at time step $t$ as
\begin{align}\label{eq:best idx}
p^\star[t] = \underset{p[t] \in \{1,\hdots,M\}}{\arg\max}\ \big|\bh^H \bff_{p[t]}\big|^2.
\end{align}
\par
In this paper, we aim at achieving accurate beam prediction and beam tracking based on LiDAR sensing information. For that, we define $\bx[t] \in \bbR^{D \times 1}$ as the LiDAR sensing signal captured by the BS at time step $t$, where $D$ represents the number of quantized angles in the LiDAR field-of-view and each entry of $\bx[t]$ represents the LiDAR-based estimated distance to the obstacle at the corresponding angular direction. The objective is then to find a mapping function which can accurately predict the optimal future beam from the sequence of observed LiDAR information denoted by $\boldsymbol{\mathcal{X}}_{t, W} = \left\{\bx[t-W+1],\hdots, \bx[t]\right\}$, where $W \in \bbZ^{+}$ is an observation window. In this paper, we focus on the use of machine learning (ML) models to learn this mapping function. Let $f(\boldsymbol{\mathcal{X}}_{t, W};\Theta)$ denote an ML model that takes $\boldsymbol{\mathcal{X}}_{t, W}$ and predict the current/future beams, with  $\Theta$ representing the ML model parameters. Given the input $\boldsymbol{\mathcal{X}}_{t, W}$, the objective is to design the ML model to accurately predict the optimal current/future BS beams as defined in~\eqref{eq:bbest beam}. Next, we formally define the beam prediction and tracking problems.
\par
\textbf{LiDAR Beam Prediction:} The beam prediction task is defined as predicting the \textit{current} optimal beam based on the LiDAR data captured until the \textit{current} time step. Mathematically, the optimal ML model for beam prediction can be written as
\begin{align}\label{eq:optm1}
f_{pr}^\star(\boldsymbol{\mathcal{X}}_{t, W};\Theta_{pr}^\star) = \underset{f_{pr}(\cdot), \Theta_{pr}}{\arg\max} \ \bbP\left\{f_{pr}\left(\boldsymbol{\mathcal{X}}_{t, W};\Theta_{pr}\right)=p^\star[t]\right\},
\end{align}
where $f_{pr}(\cdot)$ denotes the ML model designed for the beam prediction task, and $\Theta_{pr}$ is its training parameters. It is worth noting here that \eqref{eq:optm1} also leverages the LiDAR data captured at previous time steps since this data is available and only requires relatively low storage overhead.
\par
\textbf{LiDAR Beam Tracking:} The beam tracking targets predicting the optimal \textit{future} beams using the LiDAR data captured until thes current time step. Similar to the beam prediction problem, the optimal ML model $f_{tr}^\star(\cdot)$ for beam tracking can be expressed as
\begin{align}\label{eq:optm2}
f_{tr}^\star(\boldsymbol{\mathcal{X}}_{t, W};\Theta_{tr}^\star) = \underset{f_{tr}(\cdot), \Theta_{tr}}{\arg\max} \ \bbP\left\{f_{tr}\left(\boldsymbol{\mathcal{X}}_{t, W};\Theta_{tr}\right)=p^\star[t+ v]\right\},
\end{align}
where $\Theta_{tr}^\star$ is the associated trainable parameters, and $v \in \{1, ..., V\}$ denotes the lead-time (future instance) of the beam tracking process. Note that the beam prediction task in \eqref{eq:optm1} is a special case with $v=0$. Next, we present a baseline for the beam tracking problem which relies only on wireless data. 
\par
\textbf{Baseline Beam Tracking:} The beam sequence adopted to serve the UE in the previous time steps contains information about the UE moving pattern, which can be exploited to predict the future optimal beams. In this paper, we employ this baseline ML model which predicts the future beams using the previously used beams. Formally, this is defined as
\begin{align}\label{eq:optm3}
f_{b}^\star(\boldsymbol{\mathcal{F}}_{t,W};\Theta_{b}^\star) = \underset{f_{b}(\cdot),\Theta_{b}}{\arg\max} \ \bbP\left\{f_{b}\left(\boldsymbol{\mathcal{F}}_{t,W};\Theta_{b}\right)=p^\star[t+v]\right\},
\end{align}
where $\boldsymbol{\mathcal{F}}_{t,W} = \left\{ \bff[t-W+1],\hdots,\bff[t] \right\}$ denotes the  beam sequence used at time steps $(t-W+1)$, ..., $t$. Similar to the two LiDAR beam management tasks, $f_{b}^\star(\cdot)$ and $\Theta_{b}^\star$ are the optimal ML model and trainable parameters associated to this baseline beam tracking task.
\par
In Section \ref{Proposed Soulution}, we explain the proposed ML models for mmWave beam prediction and beam tracking.

\section{Proposed LiDAR Aided Beam Management Solution}\label{Proposed Soulution}
First, we present the key idea and motivation for why utilizing LiDAR data is promising for the mmWave/THz beam management problems. Then, we describe the proposed machine learning model for the LiDAR beam prediction/tracking tasks and the baseline (beam history based) solution.

\subsection{Key Idea: Why LiDAR?}
Leveraging mmWave/THz frequency bands is essential for meeting the increasing data rate demand. Accompanied by the large available bandwidth, however, communication in these bands has new challenges. In particular, one major feature of these systems is the need to use large antenna arrays and narrow beams at the transmitters and receivers to achieve sufficient receive power. Adjusting these narrow beams, however, is associated with high beam training overhead \cite{3gpp}. This  overhead affects the spectral efficiency and the latency, and becomes severer for highly mobile and dynamic scenarios.

Overcoming the beam training overhead challenges requires developing new and novel solutions. One important observation for mmWave/sub-THz communication systems is that most of the transmission power is directed towards the line-of-sight (LoS) paths. Therefore, the direction/position of the communication terminals and the geometry of the surrounding environment become important factors for determining the optimal beam. This motivates utilizing sensing information to build such awareness about the surrounding environment and transmitter/receiver locations. In this paper, we focus on leveraging the LiDAR sensory data to help guide the beam management tasks and attempt to draw insights into their potential gains and possible limitations.

\subsection{Proposed LiDAR-Aided System Operation}
We propose a LiDAR-aided mmWave communication system that operates as follows: In each time step (\textit{e.g.}, coherence time), the BS captures a LiDAR image using its sensor as shown in Fig.\ref{fig:system_model}. A sequence of $L$ captured LiDAR sensing images is then presented to an ML model, which predicts the top-$k$ promising beams that should be used to serve the user. Given the  top-$k$ predicted beams, the BS can either (i) directly adopt the top-$1$ beam and completely eliminate the beam training overhead or (ii) perform an over-the-air beam refinement/training using only the predicted subset of beams (the top-$k$ beams). Next, we describe the proposed ML model. 

\subsection{Deep Learning Model}
Recurrent neural networks (RNNs) are known for their high efficiency in sequential modeling \cite{RNN}. Since the objective of this paper is current and future beam prediction based on the observed sequence of sensory data, we utilize an RNN architecture for sequential feature extraction. Fig. \ref{fig:system_model} shows the block diagram of the proposed beam prediction and tracking ML model. As shown in \figref{fig:system_model}, the ML model consists of (i) $W-1$ repeated blocks, each has an embedding and a gated recurrent unit (GRU) \cite{gru} feature extractor components, in addition to (ii) $V+1$ repeated blocks, each  has an embedding, a gated recurrent unit (GRU) \cite{gru} feature extractor, and classifier components. Next, we describe the different components.

\textbf{Embedding Block}: The embedding block maps the raw input information into a different representation. This block has two major advantages. First, it can often compress the raw input vector into a smaller vector space, which decreases the number of trainable parameters in the successive blocks and stabilizes the training process. Second, the embedded vector captures some semantics of the raw input vector such that semantically-similar input vectors are close together in the embedding vector space. Therefore, the embedded representation is easier to extract features from. For the LiDAR data embedding, we adopt a fully connected layer that projects the $\bx[t]\in \bbR^{D \times 1}$ into $\tilde{\bx}[t]\in \bbR^{D_e\times 1}$. To embed the previous beam indices, we apply a trainable look-up table of $M$ embedding vectors $\left\{ \tilde{\bp}_1,\hdots,\tilde{\bp}_M \right\}$. Each embedding vector $\tilde{\bp}_m \in \bbR^{M_e \times 1}$ corresponds to the beam index $m\in\{1,\hdots,M\}$. In the experimental results, we set $D_e$ and $M_e$ to $64$.
\par
\textbf{RNN Feature Extractor Block:} The second component of the proposed ML model is the RNN block. We employ $W$ single-layer GRUs to extract features from the input LiDAR or beam sequence. In our implementation, the hidden state size of the GRU is $64$, and it is initialized with the all-zero vectors. For fair comparisons, we use the same RNN architecture in all three ML beam management tasks.
\par
\textbf{Classifier Block:} Following the RNN feature extractor, a fully connected layer works as the classifier. Since the beam prediction and tracking tasks are classification tasks, we employ the softmax activation function at the output of this fully connected layer. The output of the classifier is a score vector $\hat{\bs} = [s_1, \hdots, s_M]^T$. The score $s_m \in (0,1)$ corresponds to the $m$-th beam $\bff_m$ in the codebook. The beam index with the highest score indicates the predicted beam.
\par
\textbf{Learning Process:} The ML model is trained offline in a supervison way. Each data point consists of (i) the input sequence $\left\{\boldsymbol{\mathcal{X}}_{t,W}, \boldsymbol{\mathcal{Z}}\right\}$, where $\boldsymbol{\mathcal{Z}}_{t,W}$ is a padding sequence of $V$ all-zero vectors, each of dimension $D \times 1$, and (ii) the label which is the desired output sequence $\left\{\bs^\star[t],\hdots, \bs^\star[t+V] \right\}$, where $\bs^\star[t+v]$ is the one-hot representation of $p^\star[t+v]$. 

For the baseline model, each data point consists of (i) the input sequence $\left\{\boldsymbol{\mathcal{F}}_{t,W}, \boldsymbol{\mathcal{Z}}\right\}$, and (ii) the desired output sequence $\left\{ \bs^\star[t],\hdots, \bs^\star[t+V] \right\}$. To train the ML models, we adopt a cross-entropy loss function applied to the last $\gamma$ outputs among the output sequence, where $\gamma$ is a design parameter that is set to 4 in our implementation. 
\begin{table*}[t!]
\normalsize
\setlength\tabcolsep{4pt}
\caption{\label{tb:acc}Top-$k$ accuracy of the ML models with LiDAR and beam input trained on $80\%$ and evaluated on $20\%$ data.}
\centering
\begin{tabular}[b]{l
S[table-format=-1.2]
S[table-format=-1.3]
S[table-format=-1.2]
S[table-format=-1.3]
S[table-format=-1.2]
S[table-format=-1.3]
S[table-format=-2.2]
S[table-format=1.2]
S[table-format=-2.2]
S[table-format=1.2]
S[table-format=-2.2]
S[table-format=-1.2]}
\toprule
\multicolumn{1}{c}{Metric}& \multicolumn{2}{c}{Current beam} & \multicolumn{2}{c}{Future beam 1}& \multicolumn{2}{c}{Future beam 2}& \multicolumn{2}{c}{Future beam 3}\\
\cmidrule(lr){1-1} \cmidrule(r){2-3} \cmidrule(lr){4-5}\cmidrule(lr){6-7}\cmidrule(lr){8-9}
&{LiDAR input} & {Beam input} & {LiDAR input} & {Beam input} &{LiDAR input} & {Beam input} & {LiDAR input} & {Beam input}\\
Top-1 & {57.5\%} & {-} & {55.3\%} & {60.3\%} & {51.9\%} & {56.8\%} & {46.0\%} & {50.9\%}\\
Top-2 & {80.6\%} & {-} & {80.4\%} & {83.6\%} & {74.9\%} & {82.0\%} & {70.1\%} & {78.3\%}\\
Top-3 & {89.5\%} &{-} &{88.3\%} & {92.1\%} & {86.5\%} & {91.4\%} & {82.6\%} & {89.9\%}\\
Top-5 & {95.6\%} &{-} &{95.0\%} & {97.8\%} & {94.5\%} & {97.0\%} & {93.3\%} & {96.7\%}\\
\bottomrule
\end{tabular}
\end{table*}

\section{Evaluation Setup}\label{Simulation Setup}
In this section, we present the first evaluation of LiDAR aided beam prediction based on a large-scale real-world dataset. In particular, we adopt our DeepSense 6G dataset \cite{DeepSense}, which comprises co-existing multi-modal sensing and communications data. Next, we describe the adopted DeepSense scenario and the task-specific development dataset. 

\subsection{DeepSense 6G Secnerio 8}
\begin{figure}[t]
\centering
\includegraphics[width=1\linewidth]{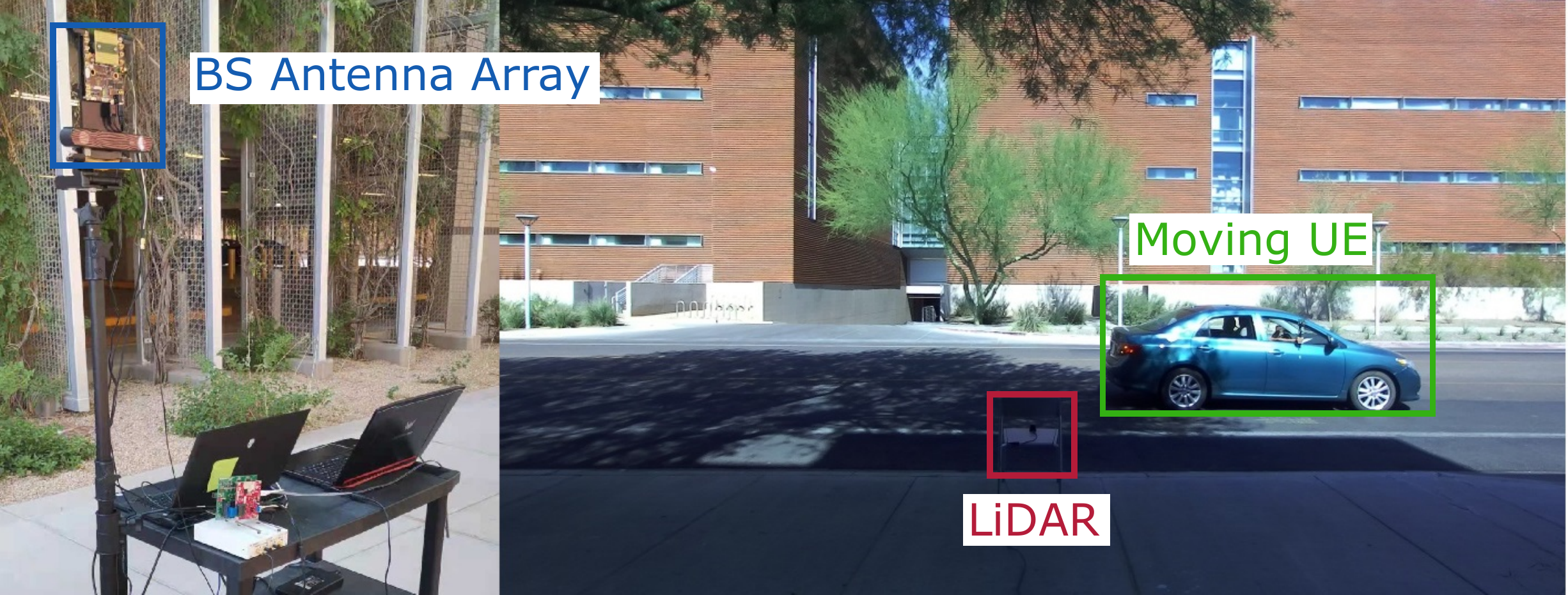}
\caption{System setup of the Secnerio 8 of the DeepSense 6G dataset. The mmWave BS antenna arrays receives the signal transmitted the moving UE. A LiDAR is placed in front of the BS to obtain sensing information.}
\label{fig:setup}
\end{figure}

We adopt Scenario 8 of the DeepSense 6G dataset. The system setup of Scenario 8 is shown by \mbox{Fig. \ref{fig:setup}}; it consists of (i) a moving UE working as a transmitter, (ii) a fixed BS working as a receiver and a LiDAR sensor. The moving vehicle carries an  omni-directional mmWave 60GHz transmitter that is communicating with the BS. The BS is equipped with a 60GHz receiver that has a 16-element phased array and uses a pre-defined beam codebook of 64 beams. The BS utilizes a LiDAR sensor placed in front of the transceiver to obtain sensing information of the communication environment. We refer the readers to \cite{DeepSense} for more details regarding the data collection. During the data collection process, and at each sampling time $t$, one LiDAR sensing data $\bx[t]$ and one 64-element beam training power vector (carrying the receive power values with the 64 beams) are collected.

\subsection{Devolopment Dataset Generation}
The DeepSense 6G Scenario 8 data consists of multiple data sequences. Each data sequence contains the data corresponding to only one pass through the road. we extract the optimal beam indices of each data sequence from the receive powers. After that, each data point in these data sequences is a tuple of LiDAR sensing data and optimal beam index $(\bx[t], p^\star[t])$. Then, $80\%$ of the data sequences are used to form the training dataset, and the remaining data sequences form the test dataset. Note that, with this data split, no overlapping data point (and the corresponding LiDAR information) exists in the training and test datasets so the split is data-leakage free. In the training process, we use an observation window \mbox{$W=8$}, and train the models to predict the beams up to the third future beam ($V=3$). Therefore, we further format the training dataset such that each training sample consists of (i) a LiDAR data sequence of eight time steps, (ii) the corresponding eight optimal beam indices, and (iii) three optimal beam indices for the three future time steps we intend to predict.

\section{Evaluation Results}\label{Simulation Results}
In this section, we evaluate the performance of the proposed LiDAR aided beam prediction/tracking solution compared to the baseline approach. The evaluation metric adopted is the top-$k$ accuracy. The top-$k$ accuracy is defined as the percentage of the test samples whose ground-truth beam index lies in the predicted beams of top-$k$ scores.
\par
\textbf{Beam Prediction and Tracking Accuracy:} Table \ref{tb:acc} presents the top-$1$, top-$2$, top-$3$, and top-$5$ accuracies of the two proposed approaches for the beam prediction/tracking tasks for various future instances. It can be seen that as the lead-time increases, the accuracy performance of both two models decreases, which is expected. We note that the LiDAR-based solution can achieve comparable beam prediction and tracking performance to the baseline solution. For example, the top-$5$ accuracy of the LiDAR beam prediction is $95.6\%$. This means that, {with the proposed LiDAR beam prediction model, the BS can find the optimal beam in 95.6\% of the cases, reducing the beam training overhead from 64 (in the exhaustive search case) to 5.} It is important to mention here that the baseline model is assumed to have perfect knowledge of the previous 8 optimal beams, which requires high beam training overhead.
\par
\begin{figure}[h!]
\centering
\includegraphics[width=1.0\linewidth]{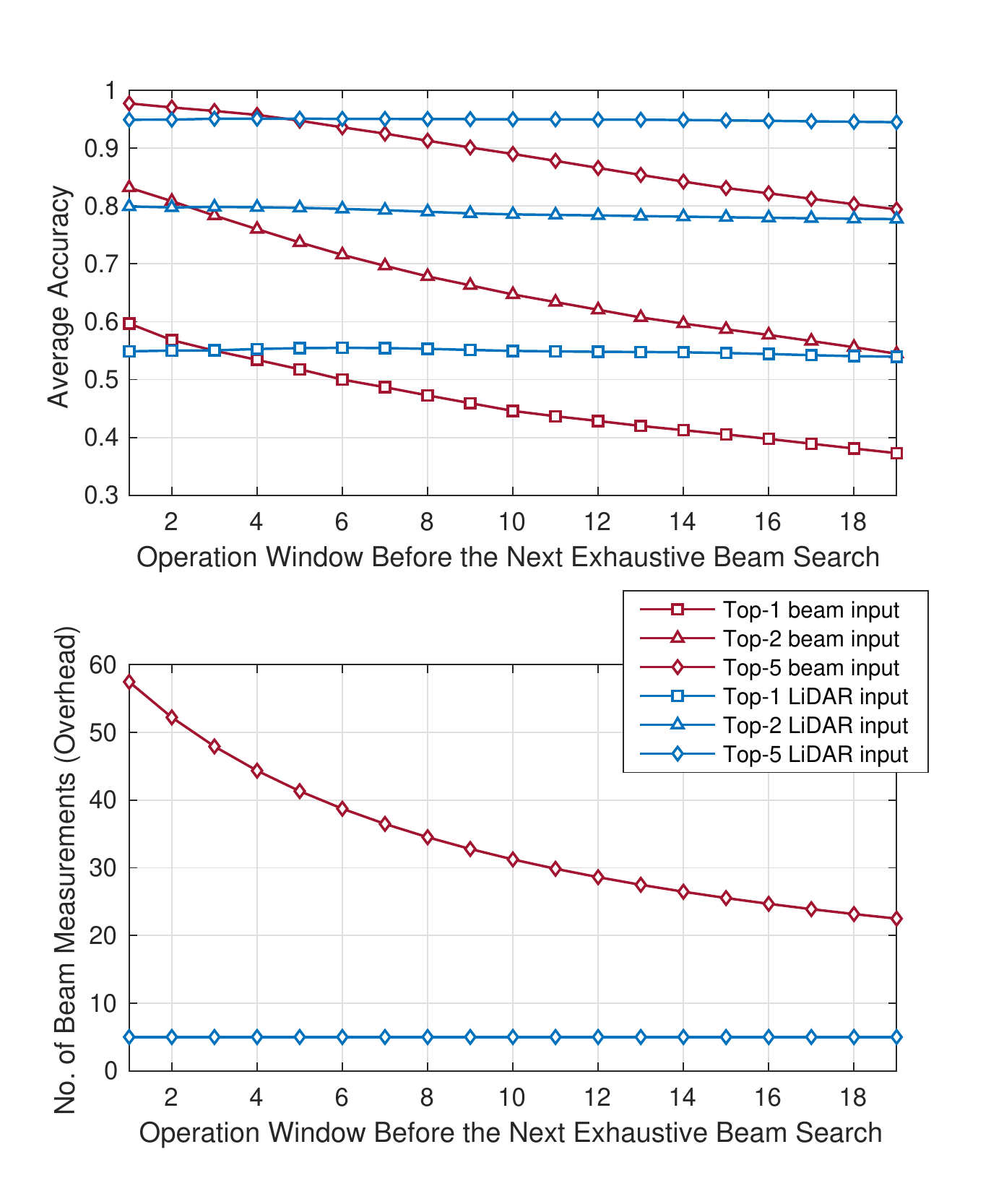}
\caption{This figure shows the performance of the two future beam prediction approaches with different operation windows. The LiDAR-aided solution can keep track of the beam with lower training overhead.}
\label{fig:longtest1}
\end{figure}

\textbf{Deviation in Future Beam Prediction:} One major advantage of the LiDAR beam prediction and tracking model is that it does not require any beam training (it eliminates or significantly reduces this overhead). This is different than the baseline beam tracking model which requires the optimal beam indices for the last $8$ time steps. Obtaining the optimal beams for the previous time steps has high beam training overhead. As an alternative solution, the baseline model can use the latest predicted beam index as the optimal index. The error in the predicted beam indices, however, accumulates and the future beam prediction is expected to gradually deviate from the ground-truth optimal beams. This deviation is captured in Fig. \ref{fig:longtest1} which presents the average \mbox{top-$k$} accuracy for predicting the \textit{first} future beam versus different \textit{operation windows}. The operation window represents the window over which the predicted beams are used as inputs to the baseline model; i.e., instead of the optimal beams, before another exhaustive beam search is done. For example, for an operation window of 5, and given that we adopt an observation window of 8, the approach predicts the first future beam  5 times (in 5 sequential time instances), and then updates its beams by performing exhaustve beam search 8 times (the length of the observation window).  It can be seen from Fig. \ref{fig:longtest1} that the LiDAR-aided approach can keep track of the beams because the BS captures the up-to-date LiDAR sensing information at every time step with no cost on the wireless resources. On the other side, the accuracy  of the baseline model degrades as the prediction window becomes longer. {Without very frequent exhaustive beam training, the proposed LiDAR aided beam prediction approach consistently outperforms the baseline solution.}
\par
\textbf{Training Overhead:} Fig. \ref{fig:longtest1} compares the average beam training overhead required for one time step versus the operation window length when using the top-$5$ beams. With a prediction window of three time steps, {the LiDAR-aided approach only needs around 10\% training overhead compared to the baseline solution.}

\section{Conclusion}\label{Conclusion}
This paper develops a machine learning based mmWave beam prediction/tracking approach utilizing LiDAR data and demonstrates its performance on a large-scale real-world dataset for the first time. Based on the DeepSense 6G dataset,  the proposed LiDAR-aided beam prediction and tracking model achieves only slightly lower accuracy than a baseline model that has perfect knowledge of the previous optimal beams. The top-$5$ accuracy of the LiDAR-aided approach is $95.6\%$ and $95.0\%$ for the current beam (beam prediction) and the first future beam (beam tracking), respectively. However, the LiDAR-aided approach only needs $10.4\%$ beam training overhead to match the performance of the baseline approach. These results highlight the potential of leveraging LiDAR sensory data in real-world mmWave communication systems.


\end{document}